\documentclass[aps, prl, 10pt, twocolumn, groupedaddress, superscriptaddress]{revtex4-2}

\usepackage[british]{babel}
\usepackage{graphicx}

\usepackage{amssymb}
\usepackage{physics} 
\usepackage{hyperref}

\newcommand{\kings}{Department of Physics, King's College London, Strand, London, WC2R 2LS, UK.}
\newcommand{\exeter}{Department of Physics and Astronomy, University of Exeter, Stocker Road, Exeter, EX4 4QL, UK.}
\newcommand{\potsdam}{Institut f\"ur Physik und Astronomie, University of Potsdam, 14476 Potsdam, Germany.}
\newcommand{\lcn}{London Centre for Nanotechnology, Department of Physics, King's College London, Strand, London, WC2R 2LS, UK.}

\newcommand{\dbar}{\delta} 
\newcommand{\tq}{{\mathrm{eq}}}
\newcommand{\supp}{Supplemental Material}
\newcommand{\ematt}{End Matter}

\begin{document}

\title{Extreme-temperature single-particle heat engine}

\author{M. Message}
\affiliation{\kings}

\author{F. Cerisola}
\affiliation{\exeter}

\author{J.D. Pritchett}
\affiliation{\kings}

\author{K. O'Flynn}
\affiliation{\kings}

\author{Y. Ren}
\affiliation{\kings}

\author{M. Rashid}
\affiliation{\kings}

\author{J. Anders}
\affiliation{\exeter}
\affiliation{\potsdam}

\author{J. Millen}
\email{james.millen@kcl.ac.uk} 
\affiliation{\kings}
\affiliation{\lcn}

\begin{abstract}

There are many exotic thermodynamic processes that are hard to study in nature. Here, we synthesize a structured environment to explore the extremes of thermodynamics. We present an engine running at extreme temperatures {of above ten Mega-Kelvin}. Our underdamped engine is realised by electrically levitating and controlling a charged microparticle in vacuum. 
Giant fluctuations are observed in the engine's heat exchange with the environment, while its efficiency shows stochastic events where more work is performed by the engine than heat consumed. Moreover, the non-uniformity of the synthetic environment leads to the  particle experiencing position dependent diffusion, a critical phenomenon in microscale biological processes.
We theoretically account for the effects of multiplicative noise and find excellent agreement with the observed behavior.

\end{abstract}

\maketitle

The thermodynamic behaviour of microscopic systems is full of surprises; engines can run backwards for a short time \cite{Blickle2012}, diffusion can be directed \cite{HanggiReview2009} and the thermal environment remembers where you were \cite{Franosch2011}. Accurate models of microscale thermodynamics are critical for understanding transport in cell-biology \cite{BressloffReview2013} and for the design of micromachines. When the fluctuation in the exchange of energy between a system and its environment becomes comparable to the energy of the system itself, we must move beyond only considering averaged behaviour, and study individual stochastic trajectories \cite{Seifert2012, Ciliberto_rev2017}. 

Single microparticles confined in harmonic potentials, typically created by optical tweezers \cite{Spesyvtseva2016}, are recognized as the paradigmatic system in which to study stochastic thermodynamics \cite{Seifert2012, Gieseler2018Review, Millen2018Review}, and come with an impressive toolbox of control techniques \cite{Spesyvtseva2016, Gieseler2021, MillenReview2020, G-Ballestero2021rev}.
This platform has enabled seminal studies of information thermodynamics \cite{Toyabe2010, Berut2012} and elucidated microscopic thermal dynamics \cite{Gomez-Solano2009, Rings2010, Franosch2011, Ibanez2024}. By levitating objects in a gas of controllable pressure, one can tune the rate at which they exchange energy with their environment, which has allowed observation of ballistic Brownian motion \cite{Li2010}, equilibration at the single-trajectory level \cite{Gieseler2014, Raynal2023}, non-equilibrium energetics \cite{Hoang2018,Debiossac2020,Militaru2021} and the transition from under- to over-damped bistability \cite{Rondin2017, Ricci2017, Militaru2021}. 

\begin{figure}[t]
\centering
\includegraphics[width=0.49\textwidth]{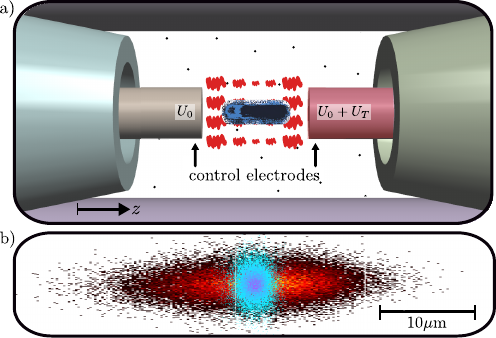}
\caption{\textbf{Schematic of the single particle engine.} \textbf{a)}
A charged silica microparticle
(image of the motion in blue/black, recorded by an event based camera~\cite{Ren2022}) is levitated within a linear Paul trap under vacuum conditions. Coaxial endcap control electrodes
provide harmonic confinement along the $z$-axis, with a frequency that can be varied by changing the voltages $U_0$. 
An additional fluctuating voltage $U_T$ with white-noise statistics applied to one control electrode generates a spatially varying synthetic heat bath (red lines, amplitude qualitatively indicates noise strength in space), with temperatures in excess of {$10^7$~K.} \textbf{b)} Probability distribution function of particle position in the $y-z$ plane for a particle in hot (red, {$T_h~=~2.7\times10^6$\,K) and cold (blue, $T_c~=~1.2\times10^5$\,K)} baths.}
\label{fig1}
\end{figure}

{In this work, we explore the underdamped thermodynamics of a single particle Stirling engine driven by the heat exchange between a hot bath over 100 times hotter than the cold environment. This unprecedented temperature contrast is much greater than seen in commercial macroscopic (1.3-2.8) or lab-based microscopic engines (2-9) \cite{Rossnagel2016, Martinez2016}. The engine is created by levitating a charged microparticle in a Paul trap under vacuum conditions, see Fig.~\ref{fig1}, and high temperature environments of over {$10^7$\,K} are synthesized through the use of noisy electric fields \cite{Martinez2016}.
Operating in the underdamped regime \cite{Dago2021,Dago2022,Dago2023,Klaers2017} enhances thermodynamic fluctuations as compared to experiments in liquid \cite{Blickle2012, Martinez2016,Krishnamurthy2016}. 
Moreover, this microscopic engine experiences multiplicative noise, leading to position-dependent diffusion. This is a key feature present in many biological micro-systems \cite{Hummer2005, Best2010, Venable2019, Whitford2013}.
Here we introduce a stochastic model accounting for this anomalous diffusion, and show how it plays a fundamental role in accurately predicting the fluctuations of the engine.


{\it Experimental set-up}. Our engine is realised by levitating a 4.82\,$\mu$m diameter spherical silica particle with a charge-to-mass ratio of {$q/m=(-0.047\pm0.001)$\,C/kg}, corresponding to a negative charge in excess of {$10^4\,e$}. Levitation is achieved electrically with a linear Paul trap formed by four cylinders arranged on the corners of a square, and two additional co-axial cylindrical endcap control electrodes separated by 1.66\,mm  positioned either side of the particle, see Fig.~\ref{fig1}. The levitated particle moves as a 3D harmonic oscillator, where the three centre-of-mass modes of oscillation are independent, even at the highest bath temperatures, see \ematt.
By adjusting the voltage $U_0$ applied to both endcap electrodes, the trap frequency  along the $z$-direction is changed cyclically between $f_1 = 341.4\pm0.1$~Hz  and $f_2 = 316.6\pm0.1$~Hz. This is equivalent to changing the volume in a macroscopic engine cycle \cite{Seifert2012}. The change between frequencies is linear with a ramp duration of 10\,s, see Fig.~\ref{fig:VariancePlot}a). Damping is caused by the particle colliding with the room temperature gas at the operating pressure of $(2.3\pm0.4)\cross\rm{10}^{-3}\,$mbar, resulting in a momentum damping rate of $\gamma_g = 2\pi\times(1.1\pm0.1)\,$Hz. This puts our engine deep in the underdamped regime $f_{1,2} \gg \gamma_g$. 

By applying white voltage noise $U_T$ to one of the control electrodes, see Fig.~\ref{fig1}, the effective centre-of-mass temperature of a single degree-of-freedom of the levitated particle can be changed \cite{Martinez2016}.
Due to the exceptionally deep electrical potential of the Paul trap ($>10^{9}\,$K), significantly exceeding the depth of commonly used optical potentials ($<10^5\,$K) \cite{MillenReview2020}, we can heat the particle in excess of {$T_\mathrm{h} = 10^7\,$K} while remaining within the linear region of the trap. This allows us to achieve temperatures far surpassing those of previous single-particle engines \cite{Martinez2016, Rossnagel2016, Li2024}.
We determine the centre-of-mass temperature of the particle via the power spectral density of its motion \cite{Millen2014}, see \ematt. Probability distributions of particle displacement can be seen in Fig.~\ref{fig1}b). At the highest temperatures, {the particle oscillates with an amplitude of $\sim50\,\mu$m.} To track this motion while maintaining both high spatial and temporal resolution we use an event based camera,  characterized elsewhere \cite{Ren2022}.
Importantly, the electric field in the $z$-direction is not uniform, see \ematt, hence the bath temperature generated by $U_T$ is not spatially uniform. Due to the amplitude of the trapped particle's motion, it experiences a position-dependent temperature.

\begin{figure}[t]
    \centering
    \includegraphics[width=0.49\textwidth]{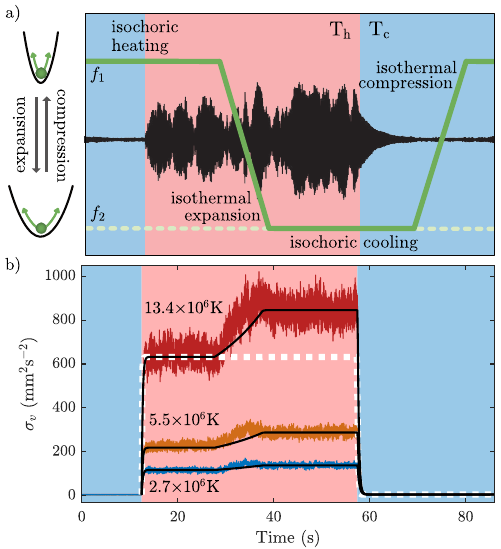}
    \caption{\textbf{Engine cycle with position dependent diffusion.}
    \textbf{a)} A measured position trajectory (black trace) of a levitated microparticle undergoing a Stirling engine cycle (green). In the isochoric heating step the temperature is changed from $T_\mathrm{c}$ (blue shaded region) to $T_\mathrm{h}$ (red shaded region). The trap frequency is subsequently changed linearly from $f_1 = 341.4 \pm 0.1$~Hz to $f_2 = 316.6 \pm 0.1$~Hz in a 10~s ramp, realizing isothermal expansion.
    \textbf{b)} Velocity variance $\sigma_v$ (coloured traces)  over 1000 cycles with a time-bin of 1\,ms.  Hot bath temperatures indicate average temperatures  $T_\mathrm{h} = (T^\mathrm{h}_1 + T^\mathrm{h}_2)/2$, where the effective temperatures at the two trap frequencies are defined as $T^\mathrm{h}_{1,2} = \frac{m}{k_\mathrm{B}} \, \sigma_v(f_{1,2})$. Note that here $\sigma_v$ denotes the variance, $\langle v^2 \rangle - \langle v \rangle^2$, not the standard deviation $\sqrt{\langle v^2 \rangle - \langle v \rangle^2}$.
    Solid black lines are obtained by numerically solving equation~\eqref{eq:fokkerplanck}, see \supp~\cite{SM}.
    The white dashed line indicates the predicted variance for standard Brownian motion with temperature $T^\mathrm{h}_{1}$ for the largest noise magnitude.
    The observed deviation of the levitated particle's velocity variance from standard Brownian motion is a clear indication of position dependent diffusion.
    }
    \label{fig:VariancePlot}
\end{figure}

We run a Stirling engine cycle as illustrated in Fig.~\ref{fig:VariancePlot}a). The particle is brought to equilibrium at a high temperature $T_\mathrm{h}$ through application of white voltage noise $U_T$. 
The Paul trap potential is quasistatically relaxed by reducing the DC voltage $U_0$, and the particle is again left to reach equilibrium.
The white voltage noise $U_T$ is switched off, and the particle thermalizes with the surrounding gas and residual voltage noise, which determines the cold temperature {$T_\mathrm{c} \approx 1.2\times10^5$ K.} An isothermal compression step is achieved by evenly increasing the voltage on both endcap electrodes, completing the Stirling cycle. This cycle is repeated 700-1400 times at each value of $T_\mathrm{h}$.


{\it Theoretical modeling}.
As we will see, a key feature of our electrically levitated particle engine is that fluctuations of thermodynamic quantities (around the mean values considered within macroscopic thermodynamics) can be enormous. Stochastic thermodynamics \cite{Seifert2012} provides the framework for the analysis of such dynamics. 
To model the dynamics of the particle, we set up the Fokker-Planck equation for the probability distribution ${\cal P}(z, v, t)$ of the particle having at time $t$ the position $z$ and velocity $v = \dot{z}$ in the $z$-direction. The voltage noise $U_T$ induces a stochastic electric field $E(z,t)=E(z)\,\xi(t)$, with magnitude $E(z)$ and $\xi(t)$ describing Gaussian white noise with zero mean, $\langle \xi(t) \rangle =0$, which is delta-correlated $\langle \xi(t) \xi(t') \rangle = 2 \delta(t-t')$, where  $\langle \cdot \rangle$ is an average over an ensemble of stochastic trajectories. 

\begin{figure*}[t]
    \centering
    \includegraphics[width=0.99\textwidth]{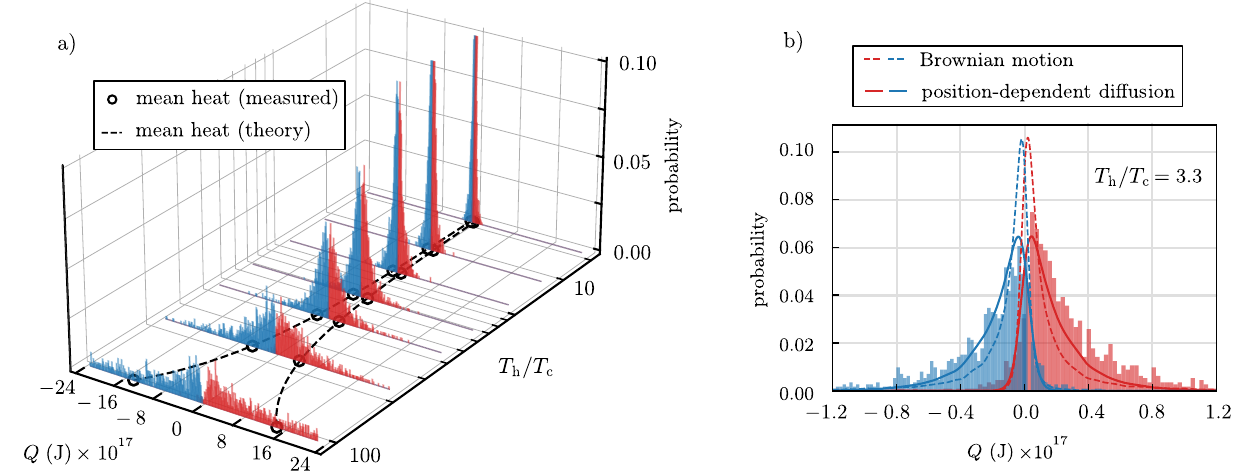}
    \caption{ {\bf Stochastic engine heat distributions.} 
    {\bf a)} Measured distribution of the heat $Q$ the levitated particle exchanges with the cold (blue) and hot (red) environment, respectively, as a function of temperature ratio $T_{\mathrm{h}}/T_{\mathrm{c}}$. The highly stochastic nature of the heat exchange is evident in the wide range of heat values. Distributions at the highest temperature ratio have been truncated for clarity.
    Negative (positive) heat values signify energy is transferred from (to) the particle's motion to (from) the bath. The mean of these experimental distributions are marked with a black circle, with the theoretical prediction (Eq.~(\ref{eq:heat_avg})) indicated by a dashed line. 
    Values of heat {reach $1.5 \times 10^{-15}$~J, equivalent to $\sim900\,k_{\mathrm{B}}T_{\mathrm{c}}$}, much larger than  previous single-particle heat engines ($<1\,k_{\mathrm{B}}T_{\mathrm{c}}$ \cite{Martinez2016,Li2024}) due to the extreme temperatures involved. 
    \textbf{b)} A pair of heat distributions at $T_{\mathrm{h}}/T_{\mathrm{c}} = 3.3$.
    The stochastic nature is particularly pronounced, as evidenced by the negative values of the particle's heat absorption from the hot bath (red), i.e. the particle sometimes dumps heat into the hot bath, and vice versa cools the cold bath (blue, positive values). 
    The experimental data is compared to predictions based on standard Brownian motion (dotted lines) and our model with position-dependent diffusion (solid lines, Eq.~\eqref{eq:heat_avg}), the latter showing better agreement with the data. 
    }
    \label{HeatDist}
\end{figure*}

Adapting the form of a general multi-variate Langevin equation~\cite{Risken1996}
to capture the experimental situation we find 
\begin{align} \label{eq:fokkerplanck}
&\partial_t {\cal P}(z, v, t) = \\    
&\left[- v \, \partial_z  + \partial_v (\omega^2 (z-z_0) + \gamma_g v) + D(z) \, \partial_v^2  \right]  \! {\cal P}(z, v, t), \nonumber
\end{align}
where $\omega = 2\pi f$ and $z_0$ are the trap angular frequency and trap centre position, respectively, which can both vary in time.
In contrast to previous microscopic engine experiments, our engine experiences position dependent noise. This is because $U_T$ generates a stochastic electric field that increases in strength when the particle approaches either electrode, as illustrated in  Fig.~\ref{fig1} (red lines). 
The noise term in Eq.~\eqref{eq:fokkerplanck} then consists of  voltage noise plus the independent noise arising from gas collisions \cite{Millen2014}, i.e. 
$D(z) =  \left(\frac{q}{m}\right)^2  E^2(z) + \frac{1}{m^2}F^2_{\rm gas}$.
Together these noises play the role
of the standard temperature term $\gamma_g k_\mathrm{B} T/m$ of Brownian motion. 

We now expand the field up to second order in $z$ since, as we will see later, the second order correction will play a crucial role in modeling the measured steady-state distributions.
The expansion $D(z) \approx D_0 + D_1(z-z_0) + D_2 (z-z_0)^2$ with $D_0 :=  \left(q/m\right)^2 E^2(z_0) + (1/m^2) F^2_{\rm gas}$, brings up  the inbuilt position-dependent diffusion terms $D_1 := \left(q/m\right)^2 \left.\partial_zE^2(z)\right|_{z=z_0}$ and $D_2 := (1/2)\left(q/m\right)^2 \left. \partial_z^2 E^2(z)\right|_{z=z_0}$.
These terms describe a bath with a position-dependent synthetic temperature $T(z) : = \frac{m}{\gamma_{g} k_\mathrm{B}} (D_0 + D_1(z-z_0) + D_2 (z-z_0)^2)$. 
This scenario experiences not only additive noise, but also multiplicative noise, see \ematt.
This makes the dynamics of our system distinctly different from standard Brownian motion and is known to give rise to a wide variety of complex phenomena in stochastic processes \cite{Volpe2016}.


{\it Results.}
We now report on the measured properties of the single-particle Stirling cycle.  Fig.~\ref{fig:VariancePlot}b) shows the time evolution of the measured velocity variance $\sigma_v = \langle v^2 \rangle - \langle v \rangle^2$ of the levitated particle at three different levels of voltage noise, corresponding to particle temperatures of {$T_\mathrm{h} \approx 2.7\times10^6$~K, $5.5\times10^6$~K, and $1.3\times10^7$~K} (blue, orange, and red lines, respectively).
The measured signal shows very good agreement with the numerical solution (black lines) of the dynamics via Eq.~\eqref{eq:fokkerplanck}. We attribute the increased variance during quasistatic relaxation to enhanced sensitivity to electronic noise when the microparticle trap frequency is changing.
The theoretical steady state variances are
\begin{equation}
    \sigma_v^\tq = \frac{D_0}{\gamma_g - D_2/\omega^2}, \quad
    \sigma_z^\tq = \frac{\sigma_v^\tq}{\omega^2}, \quad
    \langle zv \rangle^\tq = 0,
    \label{eq:tqvariances}
\end{equation}
which are all independent of the linear noise term ($D_1$).
However, $\sigma^\tq_v$ displays a dependence on the trap frequency $\omega$ via the quadratic  contribution ($D_2$) arising from the position-dependent diffusion. 
This frequency dependence is observed in the experimental data (blue, orange, red in Fig.~\ref{fig:VariancePlot}b)), and is in stark contrast to the prediction of standard Brownian motion (white dashed lines) where $D_2 = 0$.
Put differently, the presence of the frequency dependent temperatures $T^\mathrm{h}_{1} \neq T^\mathrm{h}_{2}$ implies a breaking of the standard equipartition of energy, where $\frac{m}{2} \sigma_v^\tq = \frac{k_\mathrm{B} T}{2} = \mbox{const}$, with $T$ given solely by the magnitude of the noise $U_T$. 

\medskip

Macroscopic engines, as first conceptualised by Carnot in his seminal 1824 paper \cite{Carnot1824}, run by having the working medium receive energy (in the form of heat)  from the hot reservoir ($Q_\mathrm{h}$), and dump less energy (also heat) into the cold reservoir ($Q_\mathrm{c}$), while extracting the difference in energy as useful work $W$. 
A key difference in microscopic engines is that the energetic exchanges are noticeably stochastic. Given the position dependent temperature of the particle, quantification of heat exchange can be particularly subtle.
Indeed, for the much explored overdamped case~\cite{Celani2012,Bo2013,Marino2016,Sancho2015,Polettini2013} it was found that the standard overdamped approximation can give fundamentally wrong predictions of the dissipated heat~\cite{Celani2012} when position dependent noise is present.
Here, we work in the underdamped regime. Following the well-established framework of Sekimoto~\cite{Sekimoto2010} we define heat at the single trajectory level as the work of the back-action force of the environment on to the particle.
Then the differential of heat exchanged in a single realisation~\footnote{Here $\circ$ is the product in the Stratonovich sense.} is
$
    \dbar Q = \left(-\gamma_g m v + qE(z)\xi(t)\right)\circ \mathrm{d}z(t),
$
and the average heat received by the particle~\footnote{This is the total heat exchanged between the particle and the environment at a given noise amplitude $U_T$, while not resolving the fine-grained heat flows to/from thermostats of different temperature within the environment.}  becomes
\begin{equation} \label{eq:heat_avg}
    \langle \dbar Q \rangle = \left(-\gamma_g \langle{v^2}\rangle + D_0 + D_1\langle{z}\rangle + D_2 \langle{z^2}\rangle\right) \, m \, \mathrm{d}t. 
\end{equation}
The last two new terms arise entirely due to the position-dependent diffusion.

In Fig.~\ref{HeatDist} we plot the probability distributions of the measured heat that the levitated particle exchanges with the hot and cold reservoirs.
Figure~\ref{HeatDist}a) highlights the extremely wide spread of heat exchanged with the environment, which increases dramatically with increasing temperature reaching many hundreds of $k_{\mathrm{B}}T_{\mathrm{c}}$. 
Fig.~\ref{HeatDist}b) gives an example distribution, illustrating that there is a non-zero probability that heat flows in the thermodynamically ``wrong'' direction.
This effect becomes less pronounced at higher temperatures.
We compare the measured heat distributions with the numerically obtained values from the model in Eq.~\eqref{eq:heat_avg},  with (solid line) and without (dashed line) the additional diffusion terms $D_{1,2}$, the former of which better describes the experimentally observed distribution.

\begin{figure}[t]
    \centering
    \includegraphics[width=0.49\textwidth]{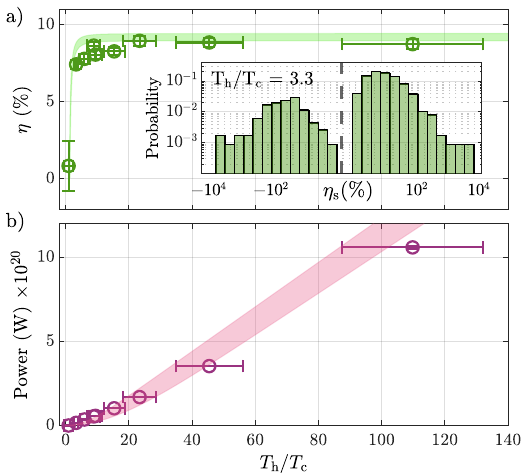}
    \caption{ 
    {\bf Engine efficiency and power over $T_{\mathrm{h}}/T_{\mathrm{c}}$.}
    {\bf a)} The measured efficiency $\eta$  (green circles) 
     agrees well with the theoretical prediction (shaded green line includes parameter uncertainties).
    Inset shows the histogram of the single-cycle efficiencies, $\eta_s = W/Q_h$, i.e. the work/heat-ratio calculated for each engine cycle realisation, at temperature ratio 3.3. The histogram highlights the highly stochastic nature of the single-cycle efficiency. Logarithmically spaced bins are used to clearly represent the full range of data, with positive and negative values separated by the grey dashed line. (Note, that $\eta = \frac{\langle W \rangle}{\langle Q_h \rangle} \neq \langle \eta_s \rangle = \langle \frac{W}{Q_h} \rangle$, due to the non-linear nature of the efficiency). Efficiency distributions with a similar bimodal shape have been discussed in Refs.~\cite{Martinez2016,Polettini2015}.
    {\bf b)} Measured average power  $P$ (purple circles) compared to theoretical curve (purple shaded region). In contrast to the saturating efficiency $\eta$, the power $P$ continues to increase with increasing temperature ratio $T_{\mathrm{h}}/T_{\mathrm{c}}$. }
    \label{fig:EfficiencyPowerPlot}
\end{figure}

From the average heat and average work, $\langle W \rangle = - \langle Q_\mathrm{h} \rangle - \langle Q_\mathrm{c} \rangle$ one can obtain the  efficiency $\eta = \langle W \rangle/\langle Q_\mathrm{h}\rangle$ and power $P = \langle W \rangle/\tau$ of the engine, where $\tau$ is the duration of the cycle. Figure~\ref{fig:EfficiencyPowerPlot}a) and b) display the experimentally derived $\eta$ and $P$, respectively, together with the theoretical predictions based on Eq.~\eqref{eq:heat_avg} (shaded areas), which indicate parameter uncertainties e.g. in particle mass and measured temperatures. 
Our levitated engine's efficiency is comparable to other underdamped engines~\cite{Li2024}, while significantly exceeding the 0.3\% achieved in the single-atom Stirling engine~\cite{Rossnagel2016}.
The maximal efficiency we obtain experimentally saturates at $9\%$, which is well below the Carnot and Curzon-Ahlborn efficiencies determined by the temperature ratio $T_\mathrm{h}/T_\mathrm{c}$. This is because the realised protocol implements a smooth change of $f$, while known optimal protocols require sudden discontinuous changes in the control parameter~\cite{Dechant2017}.
Moreover, here we explore a new regime of operation, where the particle experiences position-dependent diffusion. 
Future research will be needed to find optimal cycles in this highly non-equilibrium situation~\cite{Schmiedl2007,Dechant2017,Ye2022,Abiuso2022,Bauer2016}.

As further evidence of the highly stochastic nature of the engine, we look at the stochastic efficiency $\eta_s = W/Q_\mathrm{h}$, defined as the ratio of work and heat for each single engine cycle realisation at a given temperature ratio. This quantity has been discussed theoretically~\cite{Verley2014,Polettini2015} and experimentally~\cite{Martinez2016} elsewhere.
The inset of Fig.~\ref{fig:EfficiencyPowerPlot}a) shows the histogram of $\eta_s$ for $T_\mathrm{h}/T_\mathrm{c} = 3.3$.
We see that the efficiency distribution displays extreme values,
with efficiency fluctuations far in excess of 100\%. Moreover, in some
individual realizations the efficiency is negative. This is due to the fact that in some individual trajectories the flow of heat can be reversed and therefore the cycle is not actually operating as an engine.

{\it Discussion}.
Here we pushed Carnot's thermodynamics to the extreme by letting a single particle undergo a heat engine cycle with a synthetic bath at Mega-Kelvin temperatures. Due to the stochastic nature of this microscopic engine, there are large fluctuations in its heat exchange, including heat flowing the thermodynamically ``wrong'' way, and stochastic cycle efficiencies higher than 100\%. 
These extremes are further enhanced by operating the engine in the underdamped regime.

Additionally, this experimental platform shows great promise in its ability to
simulate and explore not only high temperatures, but also the biologically
relevant thermodynamic scenario of position-dependent diffusion (via spatially
varying temperature), which is critical in describing the dynamics of our
engine. Position-dependent diffusion is key to understanding, for example,
protein folding \cite{Best2010} and mass transport \cite{Nagai2020} in
biological settings.
Moreover, our ability to tune the system's dissipation via gas pressure would
allow studying the overdamped dynamics in the presence of position dependent
diffusion, which has attracted significant interest since it has been predicted
to show an anomalous entropy production that cannot be captured by standard
approaches~\cite{Volpe2016,Celani2012,Bo2013}.
Future research could use this platform to explore
non-Markovian energetics through the introduction of feedback
\cite{Debiossac2020, Ren2024} and thermodynamic processes in the presence of
non-white noise~\cite{Volpe2016}.
Finally, future trap designs could be engineered to make higher order
terms in the position dependent diffusion dominant, which would allow the
exploration of non-Gaussian equilibrium distributions and dynamics.

\section{Acknowledgements}
We thank Stefano Bo and Lennard Dabelow for thoughtful comments and fruitful discussions. This work has been supported by the European Union (ERC Starting Grant 803277) and the Engineering and Physical Sciences Research Council (EP/S004777/1). 
JA and FC gratefully acknowledge funding from EPSRC (EP/R045577/1).
JA thanks the Royal Society for the research grant on ``First measurements of non-equilibrium fluctuations in the underdamped regime''. JA acknowledges funding from the Deutsche Forschungsgemeinschaft (``Quantum Thermal Machines'', DFG 384846402).

\medskip 

{\it For the purpose of open access, the authors have applied a ‘Creative Commons Attribution’ (CC BY) licence to any Author Accepted Manuscript version arising from this submission.}

\medskip 

Experimental data {has been} assigned a DOI and {is} available through the KORDS research data repository \cite{KORDS}. Code to plot the simulated dynamics is available from the authors upon reasonable request, f.cerisola@exeter.ac.uk.
}
\bibliography{biblio}

\cleardoublepage 

\section{End Matter}

\begin{figure}[b]
    \centering
    \includegraphics[width=0.45\textwidth,trim={0cm 0cm 0cm 0.5cm},clip]{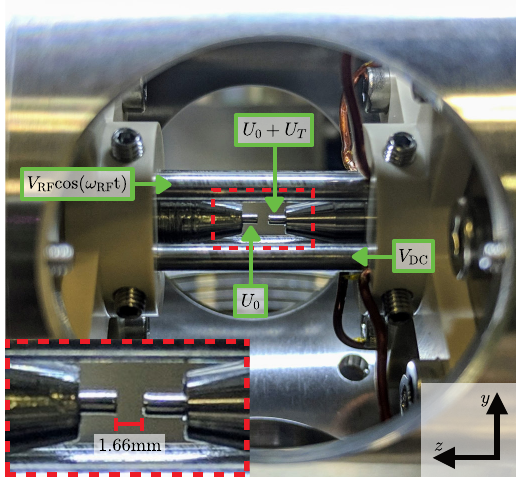}
    \caption{\textbf{Image of custom built Paul trap.} The custom built linear Paul trap sits within metal shields to protect it from stray fields. The coaxially aligned endcap control electrodes have additional shielding to minimize cross-talk between the outer trapping and the inner control electrodes.}
    \label{fig:TrapDiagram}
\end{figure}

\noindent {\it Levitation setup.~} 
A $(4.8\pm0.5)\,\mu$m diameter silica sphere (Bangs Laboratories, Inc.) is levitated at $(2.0\pm0.4)\times10^{-3}$\,mbar using a linear Paul trap, shown in Fig.~\ref{fig:TrapDiagram}. The Paul trap consists of four 3.0\,mm diameter steel rods that are positioned at the four corners of a square, with the centres of the rods on a circle of radius 5.0\,mm. A signal generator (Stanford Research Systems DS345) generates a sinusoidally varying voltage which is amplified (TREK 10/10B-HS) and applied to one pair of diagonally opposed electrodes as V~=~$V_{\rm{RF}}\rm{cos}(\omega_{\rm{RF}}t)$ where $V_{\rm{RF}}$~=~800\,V and $\omega_{\rm{RF}}$~=~$2\pi\times 1450$\,Hz. This generates a time-average harmonic potential in the $y$-$z$ plane. The other pair of diagonally opposed electrodes are used to position the particle at the centre of the potential to minimize its micromotion, and carry 0-10\,V DC.

Two 1.0\,mm diameter steel rods (endcaps) are aligned coaxially along the center of the Paul trap with a separation of 1.66\,mm. A voltage supply (Stanford Research Systems SIM928) generates a DC voltage that is amplified (Falco Systems WMA-20) to give $U_0$ = -8.0\,V on both electrodes, confining the particle in 3D.

\begin{figure}[b]
    \centering
    \includegraphics[width=0.5\textwidth]{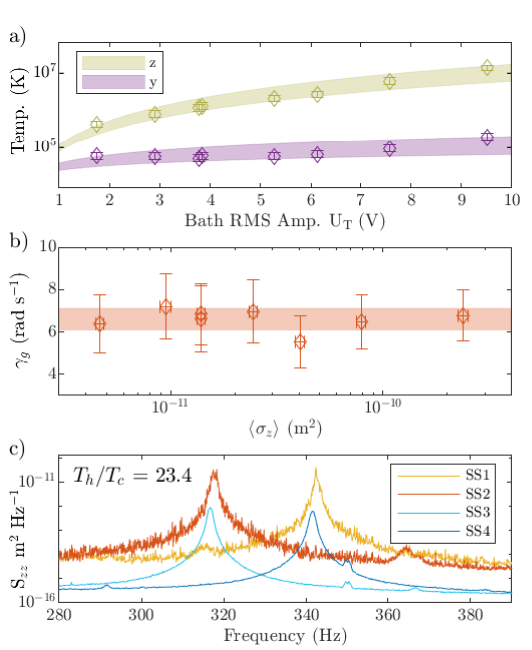}
    \caption{{\bf a)} The temperature of the particle along the $z$-axis in response to white voltage noise with RMS amplitude $U_T$ (green diamonds). The shaded region represents a power-law fit  to the data $T\propto U_T^{\alpha}$, returning $\alpha = 2.01\pm0.13$, with the fitting uncertainty reflected in the thickness of the shaded area. The noise is applied to one endcap electrode, producing a 1-D field in the $z$-direction. This is verified by simultaneously monitoring the temperature along the $y$-axis (purple diamonds). A fit to this data yields $\alpha = 0.57\pm0.15$, a significantly weaker dependence.
    {\bf b)} The measured momentum damping rate $\gamma_g$ of a particle as the variance  $\sigma_z$ of the motion along the $z$-axis increases due to increasing temperature ($T=T_h$, $f=f_1$). Since non-linearities broaden the spectral response \cite{Gieseler2013}, constant $\gamma_g$ verifies that the dynamics remain linear at all temperatures. The shaded area represents the mean of $\gamma_g$ $\pm$1 standard deviation {\bf c)} The PSD of the particle motion along the $z$-axis at all four steady states (SS) of the engine cycle. There is a small DC offset to the white noise signal which we use to heat the particle. This offset displaces the particle's mean position when the white noise is applied, leading to a temperature-dependent frequency shift of $\le 1\,\%$.}
    \label{fig:SupplamentaryPlot}
\end{figure}

The particle is trapped using Laser Induced Acoustic Desorption \cite{Bykov2019,Nikkhou2021} at a pressure of $4\times10^{-2}\,$mbar. A dry sample of microparticles is sonicated for 30 minutes, and subsequently spread onto a 0.4\,mm thick aluminum sheet. A second sheet of aluminum is placed on top and rubbed across the sample, positively charging the particles in excess of {$10^4\,e$}. This method and the mass-selectivity of the Paul trap leads to trapping of single spheres, as confirmed by light scattering.
A 532 nm laser beam (Vortran Stradus) of 25\,mW power and a beam waist radius of $\sim100\,\mu$m is used to image the particle, which scatters light onto the sensor of an Event Based Camera (EBC), (Prophessee EVK3 Gen4.1), which tracks the particle in real time. This allows us to track the particle over hundreds of micrometers while retaining a position resolution of $30\,$nm\,Hz$^{-1/2}$.
Calibration of our imaging system, calculation of particle charge and use of the EBC is described in detail in ref. \cite{Ren2022}. Once the system has been calibrated, the temperature of the particle can be calculated by analyzing the power spectral density or the position variance \cite{MillenReview2020}.

\noindent {\it Experimental engine cycle.~} 
The Stirling engine protocol is generated by an amplified (Stanford Research Systems SIM980) function generator (Moku:Lab), which generates a voltage added to both endcap electrodes resulting in a modified $U_0$. This changes the stiffness of the Paul trap, shifting all three of the particle's motional frequencies. We can only vary $f$ in the $z-$direction by $\sim25\,$Hz otherwise the frequencies in the $x,y$ directions cross $50\,$Hz (UK mains frequency) or its harmonics, causing instability. This change represents over 30 line-widths.

The same function generator produces a signal which is used to both amplitude modulate white noise (produced by a signal generator Stanford Research Systems DS345) and trigger the event based camera, which records the time that the temperature is changed. This white noise with RMS amplitude $U_T$ is added to $U_0$ and applied to a single endcap electrode. The scaling of the particle temperature with $U_T$ is shown in Fig.~\ref{fig:SupplamentaryPlot}a). This figure also shows that we only significantly heat the particle in the $z$-direction. We confirm that the particle remains in the linear part of the Paul trap potential by observing no change in the damping rate $\gamma_g$ with increasing temperature, Fig.~\ref{fig:SupplamentaryPlot}b). Furthermore, simulation in the ion-optics software SIMION verifies that our potential is harmonic over several hundred micrometers. The response of the particle in frequency space at each of the four steady states in the engine cycle is shown in Fig. \ref{fig:SupplamentaryPlot}c).

When the white noise voltage is turned off, the particle loses energy through interactions with the residual gas in the vacuum system. However, the equilibrium temperature when $U_T=0$ is approximately {$1.2\times10^5$~K}, due to amplified voltage noise in the Paul trap.

\begin{figure}[b]
    \centering
    \includegraphics[width=0.98\linewidth]{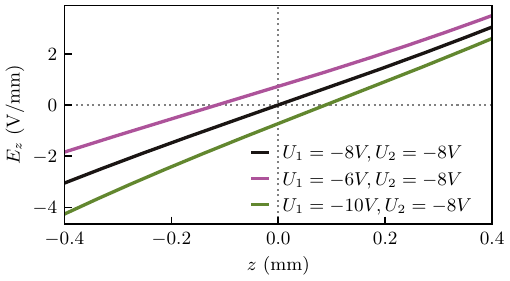}
    \caption{ \textbf{Electric field dependence.} Value of the $z$ component of the electric field given by Eq.\eqref{eq:Efield} generated by the two endcaps along the $z$-axis for three different situations: both endcaps at the same voltage (black), and the left endcap with a higher (purple) and lower (green) voltage than the right one. Note that the electric force $F_\mathrm{el} = qE(z)$ is restoring since $q < 0$.}
    \label{fig:sm-efield}
\end{figure}

\medskip

\textit{Position dependent diffusion.~} 
To understand the origin of the position dependent diffusion, we here model in
more detail the electric field applied to the particle. To simplify the system,
we will consider only the effect of the two endcaps, which confine the particle
in the $z-$axis, set its oscillation frequency~\cite{Ren2022}, and are further
used to apply the stochastic noise. We approximate each endcap by a
uniformly charged disk of radius $R = 1$~mm, so that the electric field at a
distance $z$ from a single endcap is
\begin{equation} \label{eq:Efield}
    E_\mathrm{endcap}(z) = 2\pi k\sigma\;\mathrm{sign}(z)\left(1 - \frac{|z|}{\sqrt{z^2 + R^2}}\right),
\end{equation}
where $\sigma$ is the charge density on the disk. In Fig.~\ref{fig:sm-efield} we
plot the total electric field when both electrodes are held at the same voltage
$U_1 = U_2 = U_0 = -8$~V (black line) over the range of positions explored by
the particle during a typical experiment. We see that the field is approximately
linear (and hence the potential quadratic).
The stochastic noise is applied by changing the voltage on the left electrode,
so that $U_1 = U_0 + U_T$. In Fig.~\ref{fig:sm-efield} we also plot the electric
field for the extreme cases of maximum and minimum voltage (green and purple)
for the noise level corresponding to a temperature of {$2.7\times10^6$~K} in the main text.
Again, the total field is approximately linear, $E(z) \approx E_0 + E_1z$. The
effect of the noise has two contributions: (i) it modifies the constant offset
field, $E_0$, which provides the standard thermal-like forcing, and (ii) it
modifies the linear slope, $E_1$, which determines the harmonic restoring force
and gives rise to the position dependent diffusion.




\section{Derivation of the Fokker-Planck equation}

Following the explanation in the End Matter of the main text regarding the
origin of the position dependent noise that emerges due to fluctuations in the electric
field, we describe the dynamics of the levitated particle via the Langevin
equation
\begin{equation}
    \ddot{z} = -\omega^2z -\gamma_g v + \frac{1}{m}F_\mathrm{gas}\xi_g(t) + \frac{q}{m}(E_0 + E_1z) \, \xi_e(t),
    \label{eq:app:langevinelec}
\end{equation}
where $F_\mathrm{gas}$ is the strength of stochastic forcing due to collisions
with the gas and $\xi_g(t)$ ($\xi_e(t)$) is the Gaussian white noise associated
with the gas (electric) noise such that $\langle \xi_g(t) \xi_g(t') \rangle
= \langle \xi_e(t)\xi_e(t') \rangle = 2 \delta(t-t')$ and $\langle \xi_g(t)\xi_e(t') = 0$.
From this Langevin equation, we derive a corresponding Fokker-Planck equation
for the probability density following~\cite{Risken1996}, where it is shown that
for a general set of $N$-variable ($\vec{\eta}$) Langevin equations 
\begin{equation} \label{eq:app:genericlangevin}
    \dot{\eta}_i = h_i(\vec{\eta},t) + g_{ij}(\vec{\eta},t) \,\xi_j(t),
    \qquad i = 1,\dots,N,
\end{equation}
the associated probability distribution $\mathcal{P}(\vec{\eta},t)$
follows the Fokker-Planck equation
\begin{align} \label{eq:genericfp}
    \frac{\partial\mathcal{P}(\vec{\eta},t)}{\partial t}
    &= L^\dagger_\mathrm{FP} \mathcal{P}(\vec{\eta},t), \\
  \mbox{with} \quad L^\dagger_\mathrm{FP}
    &= -\frac{\partial}{\partial\eta_i} D^{(1)}_i(\vec{\eta},t)
    + \frac{\partial^2}{\partial\eta_i\eta_j} D^{(2)}_{ij}(\vec{\eta},t),
    \nonumber
\end{align}
where the drift ($D^{(1)}_i$) and diffusion ($D^{(2)}_{ij}$) coefficients are
\begin{align}
    D^{(1)}_i(\vec{\eta},t)
    &= h_i(\vec{\eta},t)
    + g_{kj}(\vec{\eta},t)\frac{\partial}{\partial\eta_k}g_{ij}(\vec{\eta},t), \label{eq:app:genericdrift} \\
    D^{(2)}_{ij}(\vec{\eta},t)
    &= g_{ik}(\vec{\eta},t)g_{jk}(\vec{\eta},t).
\label{eq:genericdiffusion}
\end{align}
Putting our Eq.~\eqref{eq:app:langevinelec} into the form of
Eq.~\eqref{eq:app:genericlangevin}, we identify
\begin{align}
    \eta_1 = x, &\qquad \eta_2 = v, \\
    h_1(z,v,t) = v, &\qquad h_2(z,v,t) = -\omega^2z - \gamma_g v, \\
    g_{11}(z,v,t) = 0, &\qquad g_{12}(z,v,t) = 0, \\
    g_{21}(z,v,t) = \frac{1}{m}F_\mathrm{gas}, &\qquad
    g_{22}(z,v,t) = \frac{q}{m}E(z).
\end{align}
Therefore, the corresponding drift coefficients are
\begin{equation} \label{eq:app:driftcoefs}
    D_1^{(1)}(z,v,t) = v, \quad
    D_2^{(1)}(z,v,t) = -\omega^2z - \gamma_g v,
\end{equation}
and the diffusion coefficients are
\begin{align} \label{eq:app:diffusioncoefs}
    D^{(2)}_{11}(z,v,t) &= 0, \quad
    D^{(2)}_{12}(z,v,t) = 0, \quad \\
    D^{(2)}_{21}(z,v,t) &= 0, \quad
    D^{(2)}_{22}(z,v,t) = \frac{1}{m^2}F_\mathrm{gas}^2 + \frac{q^2}{m^2}E^2(z).
    \nonumber
\end{align}
Plugging this into \eqref{eq:genericfp} we obtain the
Fokker-Planck equation 
(1) of the main text. 
The diffusion coefficient is
\begin{align} 
D^{(2)}_{22}&(z,v,t) = D_0 + D_1z + D_2z^2, \quad \text{with} \\
D_0 &= \frac{1}{m^2}(F_\mathrm{gas}^2 + q^2E_0^2), \;
D_1 = 2\frac{q^2}{m^2}E_0E_1, \;
D_2 = \frac{q^2}{m^2}E_1^2. \nonumber
\end{align}
If the electric force term dominates over the gas collisions, i.e.
$F_\mathrm{gas}^2 \ll q^2E_0^2$, (which is valid due to the very low pressure of
the system), then $D_0 \approx (q/m)^2E_0^2$ and the $D_1$ coefficient is
fixed by the value of $D_0$ and $D_2$,
\mbox{$D_1 \approx 2\sqrt{D_0D_2}$}.
This removes one system parameter.

\medskip

\section{Equations of motion}

Taking the average over the ensemble of stochastic trajectories and integrating by parts the Fokker-Planck equation, we can obtain a
closed set of differential equations for the dynamics of the particle's first and
second moments, $\langle z \rangle$, $\langle v \rangle$, and $\langle z^2
\rangle$, $\langle v^2 \rangle$, $\langle zv \rangle$, which read
\begin{align}
    \frac{\mathrm{d}}{\mathrm{d}t}{\langle z \rangle} &= \langle v \rangle, \nonumber \\
    \frac{\mathrm{d}}{\mathrm{d}t}{\langle v \rangle} &= -\omega^2\langle z
    \rangle - \gamma_g \langle v \rangle, \label{eq:noneqdyn} \\
    \frac{\mathrm{d}}{\mathrm{d}t} {\langle z^2 \rangle} &= 2 \langle zv
    \rangle, \nonumber \\
    \frac{\mathrm{d}}{\mathrm{d}t} \langle v^2 \rangle &= -2\omega^2\langle zv
    \rangle -2\gamma_g\langle v^2 \rangle + 2D_0 + 2D_1\langle z \rangle +
    2D_2\langle z^2 \rangle, \nonumber \\
    \frac{\mathrm{d}}{\mathrm{d}t} \langle zv \rangle &= \langle v^2 \rangle -
    \omega^2 \langle z^2 \rangle - \gamma_g \langle zv \rangle. \nonumber
\end{align}
Note that the equations of motion for the means follow exactly the dynamics of a
standard damped harmonic oscillator, while the equations for the second moments
differ from those of standard Brownian motion due to the appearance of the
anomalous diffusion terms $D_1$ and $D_2$.
It is worth noting that we obtain a closed system of equations for just the two first moments since we keep ourselves to an expansion to second order in the noise position dependence. If higher orders became dominant, then the dynamics would become non-Gaussian and in general it would be impossible to find a closed finite set of differential equations describing the dynamics.

\medskip

Finally, setting all the time derivatives to zero, one directly recovers the steady state equations (2) of the main text.

\end{document}